\documentclass[conference]{IEEEtran}
\usepackage{flushend}

\usepackage{ifpdf}

\usepackage{cite}

\ifCLASSINFOpdf
  \usepackage[pdftex]{graphicx}
\else
  \usepackage[dvips]{graphicx}
  \DeclareGraphicsExtensions{.eps}
\fi
\begin{document}
%
\title{Simulation of Electron Beam Dynamics in a Nonmagnetized High-Current Vacuum Diode}

\author{\IEEEauthorblockN{S.Anishchenko and A.Gurinovich}
\IEEEauthorblockA{Research Institute for Nuclear Problems, Belarusian State University, Minsk, Belarus\\
Ryazan State RadioEngineering University, 59/1 Gagarina Street, Ryazan 390005, Russia\\
Email: gur@inp.bsu.by}}


%


\maketitle

\begin{abstract}
The electron beam dynamics in a nonmagnetized high-current vacuum
diode is analyzed for different cathode-anode gap geometries.
The conditions enabling to achieve
the minimal {initial} momentum spread in the electron beam are
found out.
A drastic rise of current density in a vacuum diode with a
ring-type cathode is described.  The effect is shown to be caused
by electrostatic
repulsion.
\end{abstract}

{\keywords high-current electron beam, momentum spread,
nonmagnetized vacuum diode}

%
\IEEEpeerreviewmaketitle

\section{Introduction}
%

The idea to simulate the electron beam dynamics in the
cathode-anode gap stems from the need to elucidate some of the
results obtained in the experimental study of an axial resonant
cavity vircator \cite{1EAPPC2014,TPS2015} and to achieve a better
understanding of the simulation results for various vircator
configurations \cite{4LANL,2LANL,Pasha+ZhenyaTPS}.

First, it was discovered that radiated power revealed a very
strong dependence on the cathode geometry: the maximum radiated
power was observed in the experiments with a cathode, for which
emission was confined to an annular ring on a flat cathode surface
(ring-type cathode) with a certain inner-to-outer diameter ratio
\cite{1EAPPC2014,TPS2015}. Experiments with the cathode of the
same shape and size, but with solid emission surface (solid
cathode) carried out in similar conditions gave the twice smaller
radiated power.
Similar results were obtained by the authors of
\cite{1,Korea1,SWEDEN,Israel,SWEDEN2} who reported the sensitivity
of radiated power to  cathode material, shape, and cathode-anode
gap.

Unfortunately,  despite the apparently identical experimental
conditions we failed to obtain a stable maximum-power level
signal.

In addition, in the experiments with a ring-type cathode,  an
extremely fast  (after 3-4 shots) anode mesh damage in the center
was observed, whereas for  solid cathodes such effect was absent.

%
 The most recent experimental and theoretical research of ring-type cathodes revealed the effect of  high-current  electron beam cumulation \cite{cumulation}.

%
%

Another factor that motivated the study of the beam behavior in
the cathode-anode gap was necessity to specify electron beam
parameters for XOOPIC simulation of axial vircator behavior
(XOOPIC simulates the electron beam behavior just behind  the
anode mesh).

Thus,
 to set the initial  parameters of the beam for
further simulation, one should know the parameters of the beam
leaving the cathode-anode gap. For this reason,  such parameters
should be either determined experimentally or calculated
 by simulating the electron beam dynamics in the cathode-anode gap.
Finally, the initial (unrelated to the oscillations of the
electrons reflected from the virtual cathode) momentum spread of
the electron beam was a subject of study.
(According to several publications
\cite{Bridges,Sullivan,Pasha+ZhenyaTPS} this type of  momentum
spread could also affect the operation efficiency of the
vircator.)
From this standpoint the electron beam dynamics in a vacuum diode
is discussed in present paper.
 Explosive emission is simulated with a slightly
modified hybrid code \cite{3,3a}, computing the electron and
plasma motion in different approximations: beam dynamics -- in
kinetic and inhomogeneous plasma expansion - in hydrodynamic,  thus decreasing
the runtime noticeably. Particularly, the paper analyzes the
electron beam parameters, depending on the cathode-anode gap
geometry. The conditions providing the minimal momentum
spread in the electron beam are obtained along with the
distributions of current density over the beam radius.

\begin{figure}
   \centerline{\includegraphics[width=8 cm]{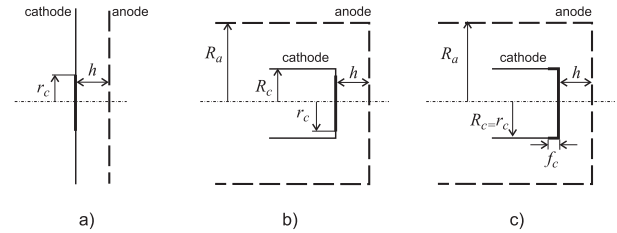}}
  \caption{Cathode-anode gap geometries}\label{fig:geometry}
\end{figure}

A 2D simulation (the system is assumed to be axially symmetric) is
carried out for three cases (Fig.\ref{fig:geometry}):
\begin{description}
    \item[-] infinite plane anode and infinite cathode with emitting
surface of radius $r_c$
    \item[-] fixed-size plane anode (of radius $R_a$) and variable size
plane cathode  (of radius $R_c$) with emitting surface of radius
$r_c$
    \item[-] fixed-size
plane anode and variable size shaped cathode
\end{description}

\section{Momentum spread}

Motion of electrons in a cathode-anode gap of nonmagnetized vacuum
diode is determined by accelerating field applied to the gap,
electrostatic repulsion and electron beam compression by
self-induced magnetic field.
Contributions from the last two factors depend on gap geometry.
Simultaneously they both also determine initial momentum spread in
the diode.
At the conducting surface of the cathode and anode electric field
has zero tangential component. When the cathode (anode) size
significantly exceeds the gap value the electron beam compression
mainly contributes to the radial motion (and beam momentum spread)
of beam electrons while the space-charge repulsion is low.
Electrons emitted near the cathode edge get into the locally
nonuniform accelerating field and undergo stronger repulsion than
those by an axis.
When the cathode (anode) size is close to or smaller than the gap
value the repulsion effect on the radial velocity (and beam
momentum spread) emerges in the foreground.
As the two factors give opposite response on momentum spread when
the cathode (anode) size is changed, the diode geometry
enabling to achieve
the minimal initial momentum spread in the electron beam is
expected to exist.
The following analysis is made to find out these conditions.

The analysis is made at fixed values of cathode-anode distance $h$; only the radius $r_c$
of the emitting surface is varied.
In cases b and c in Fig.\ref{fig:geometry}, the anode radius $R_a$
is kept fixed for these simulations and exceeds the value of $h$
ten times.
To aid in comparison of different geometries, all the results are
presented in terms of $r_c/h$ ratio.

The momentum spread
\begin{equation}
\Delta p_z / p_z=\frac{\sqrt{|\frac{1}{N_\alpha}\sum_\alpha p_{z\alpha}^2-(\frac{1}{N_\alpha}\sum_\alpha p_{z\alpha})^2|}}{\frac{1}{N_\alpha}\sum_\alpha p_{z\alpha}}
\end{equation}
is analyzed at a distance
$h/15=0.1$~cm before the anode mesh and at  constant voltage
values applied to the cathode:  256~kV and 511~kV.
%
%
The results are compared for 3 cases:

 \noindent (a)  infinite plane anode and
infinite cathode with emitting surface of radius $r_c$
(Fig.\ref{fig:geometry}a);

\noindent (b1) fixed-size plane anode and plane cathode of radius
$R_c$ equal to fixed-size radius of the emitting surface, $r_c$
(Fig.\ref{fig:geometry}b);

\noindent (b2) fixed-size plane anode  and plane cathode  whose
radius
 $R_c > r_c$ (Fig.\ref{fig:geometry}b).

\begin{figure}
   \centerline{\includegraphics[width=6 cm]{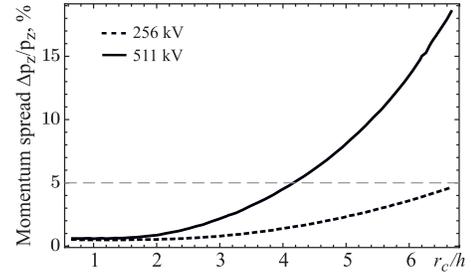}}
  \caption{Momentum spread $\Delta p_z / p_z$ for case (a) at $h=1.5$~cm}\label{fig:1_spread}
\end{figure}

In the presence of an electric field in the cathode anode gap, the
increase in  $r_c/h$ results in growth of the emitting current and
the momentum spread (Fig.\ref{fig:1_spread}). The momentum spread
is larger the higher cathode voltage is (the same result is
obtained in \cite{2}).

\begin{figure}
   \centerline{\includegraphics[width=6 cm]{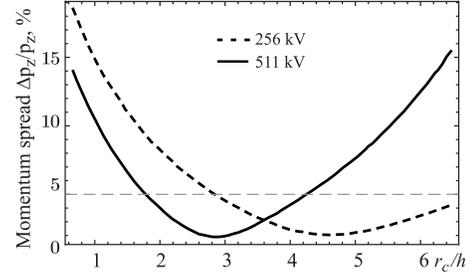}}
  \caption{Momentum spread $\Delta p_z / p_z$ for case (b1) at $h=1.5$~cm and $R_a=15$~cm}
  \label{fig:2_spread}
\end{figure}

The influence of field distortion at the cathode edges is evident
when collating Fig.\ref{fig:1_spread} with Fig.\ref{fig:2_spread},
showing the momentum spread $\Delta p_z / p_z$ as a function of
$r_c/h$.

For a plane cathode of radius $R_c=r_c$,  at
 $R_c < 2h$ the momentum spread  grows with $R_c/h$ decreasing.
The $R_c/h$ value corresponding to the momentum spread minimum
depends on the applied voltage.

\begin{figure}
   \centerline{\includegraphics[width=6 cm]{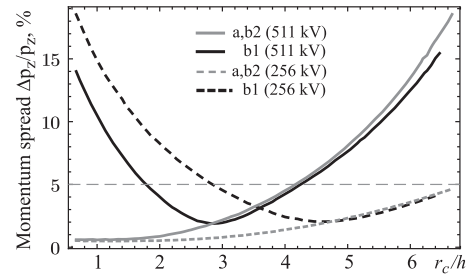}}
  \caption{Comparison of momentum spread for cases (a), (b1), and (b2) at $h=1.5$~cm, $R_a=15$~cm, and $R_c-r_c=1$~cm.}
  \label{fig:123_spread}
\end{figure}

The analysis of case (b2) gives a recipe for reducing the momentum
spread  at small $r_c/h$ ratios for a finite-size cathode: an
increase in the cathode brim (the difference $(R_c-r_c)$) enables
the transition from case (b1) to case (a).
In the considered example (Fig.\ref{fig:123_spread}), the momentum
spread for a finite-size cathode at $h=1.5$~cm, $R_a=15$~cm, and
$R_c-r_c \sim 1$~cm coincides with that for infinite-size cathode.

Being rather conceptual, the above investigation, however, gives
explicit guidelines on how to match the sizes of the cathode
(cathode brim), the emitting area and the cathode-anode gap {so as
to provide } the lowermost momentum spread in the produced
electron beam.

\begin{figure}
  \centerline{\includegraphics[width=6 cm]{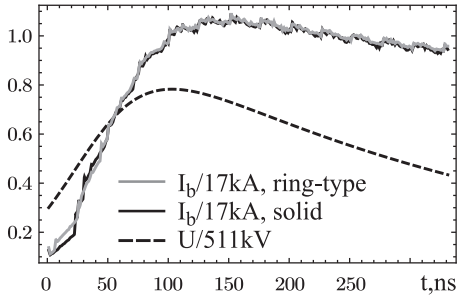}}
  \caption{Normalized beam current for the ring-type
  and solid cathodes and voltage pulse (\ref{voltage}) normalized to 511~kV}
  \label{fig:45_current+voltage}
\end{figure}

When describing a real experiment, one should consider the shape
of voltage pulse, plasma expansion and emission from the cathode
edges (see Fig.\ref{fig:geometry}c).
For further simulation, the voltage pulse (\ref{voltage}) with a
maximal value of 400~kV is used
\begin{equation}
U(t)=4 \cdot 10^5 \frac{ (1+\alpha) \alpha
^{-1+\frac{1}{1+\alpha}}}{e^{-\alpha
\frac{t}{t_p}}+e^{\frac{t}{t_p}}}, \label{voltage}
\end{equation}
 where $t_p=330$~ns and $\alpha=10$.
 The plasma expansion speed is supposed to be 2~cm/$\mu$s,
 the electric field strength sufficient for explosive emission start is 150
~kV/cm, $R_a=15$~cm, and $h=1.5$~cm.
Solid and ring-type cathodes are considered to have the outer
radii $r_c=R_c=3$~cm; the inner cathode radius for the ring-type
cathode is $r_{c1}=1.3$~cm.
The beam current normalized to $mc^3/e=17$~kA for the
ring-type  and solid cathodes and the voltage pulse
(\ref{voltage}) normalized to 511~kV are shown in Fig.
\ref{fig:45_current+voltage}.

\begin{figure}
  \centerline{\includegraphics[width=6 cm]{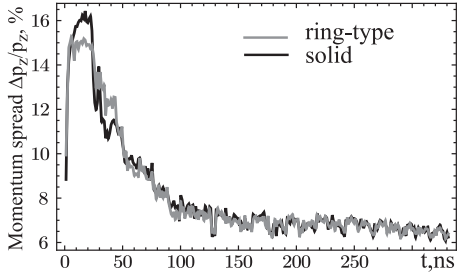}}
  \caption{Momentum spread $\Delta p_z/p_z$ for solid and ring-type cathodes}\label{fig:45_spread}
\end{figure}

For both cathode types the produced beam current (beam impedance)
is solely determined by the outer radius of the emitting surface.
While the momentum spread $\Delta p_z/p_z$ for the ring-type
cathode initially appears to be a little bit lower as compared to
that for the solid one, but in a short time ($<$ 50 ns) the
difference diminishes (see Fig. \ref{fig:45_spread}).
It should  also be mentioned that for the considered geometry
(Fig. \ref{fig:geometry}c), the momentum spread is rather high.
Perhaps, the cathode brim could reduce the momentum spread.

\begin{figure}[ht]
\begin{center}
\resizebox{45mm}{!}{\includegraphics{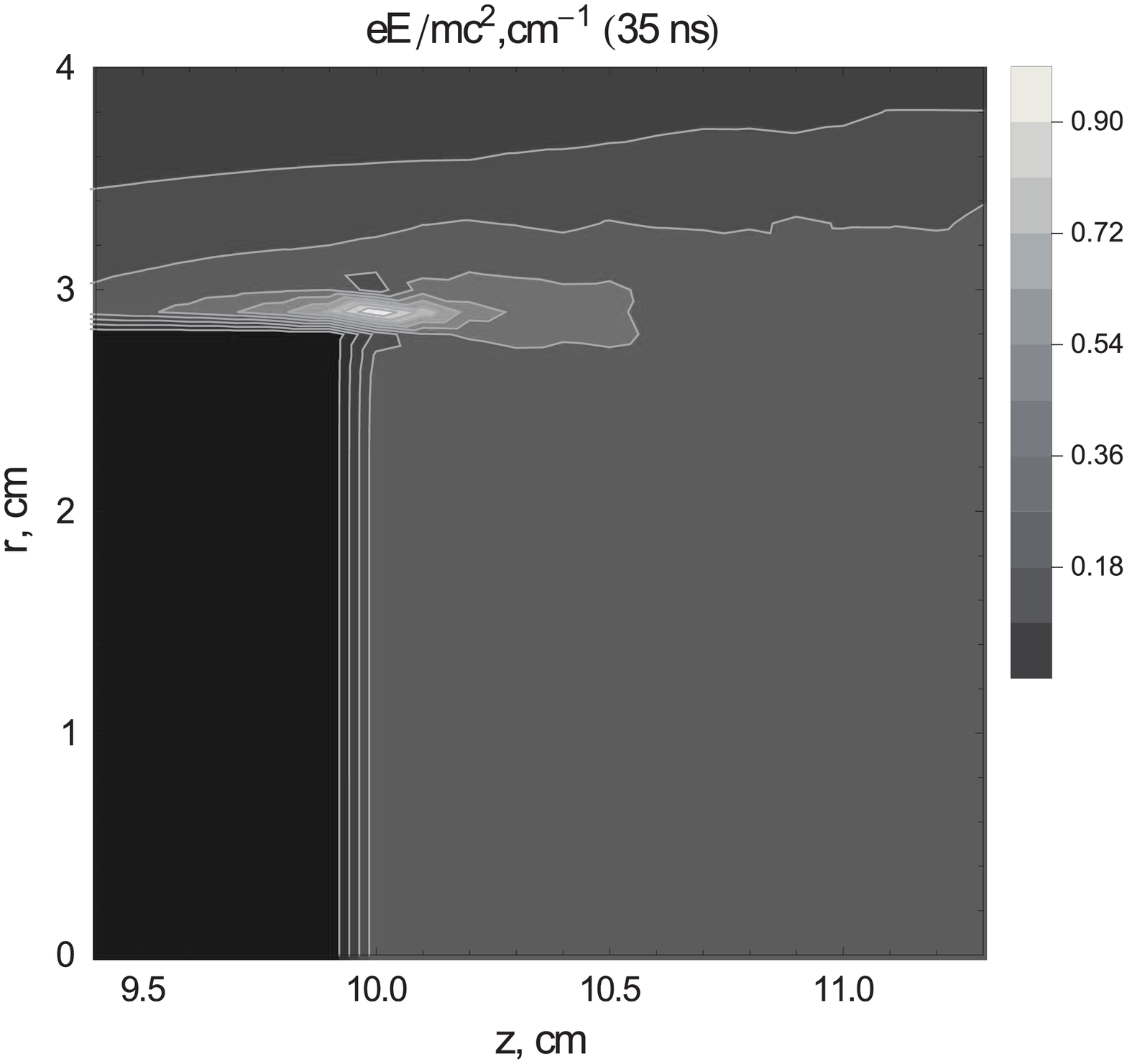}} \\ \resizebox{45mm}{!}{\includegraphics{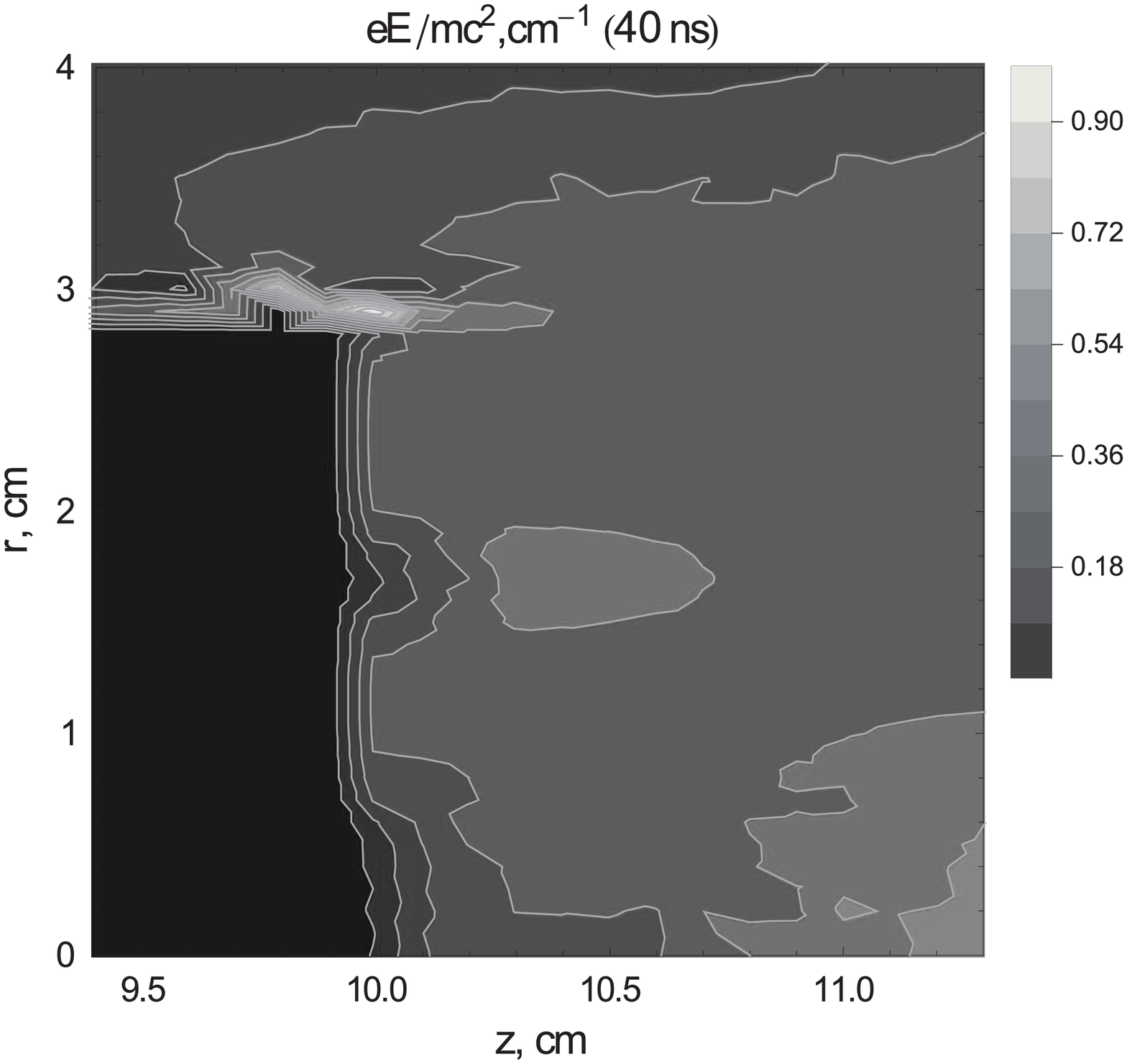}}\\
\resizebox{45mm}{!}{\includegraphics{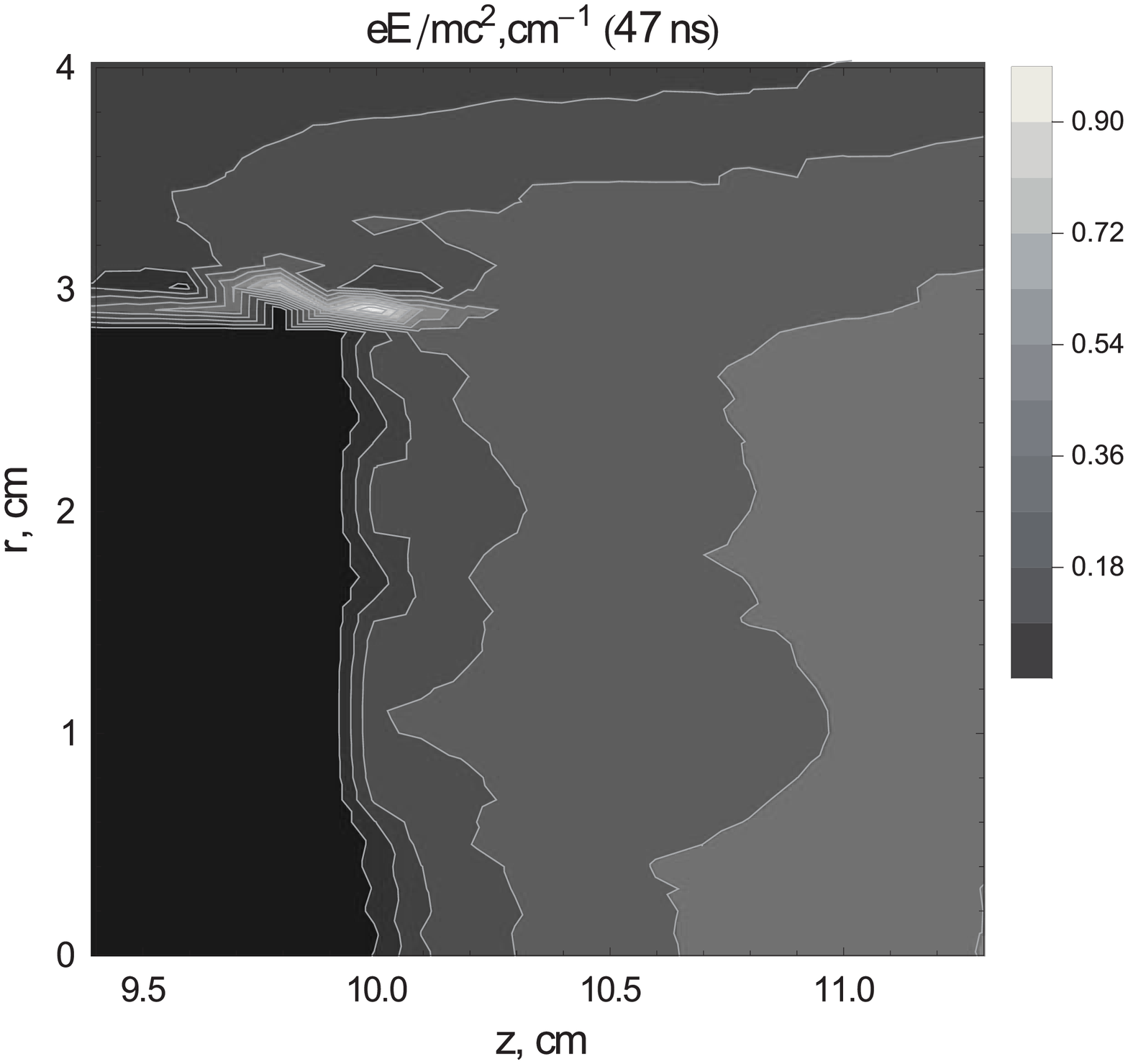}}\\ \resizebox{45mm}{!}{\includegraphics{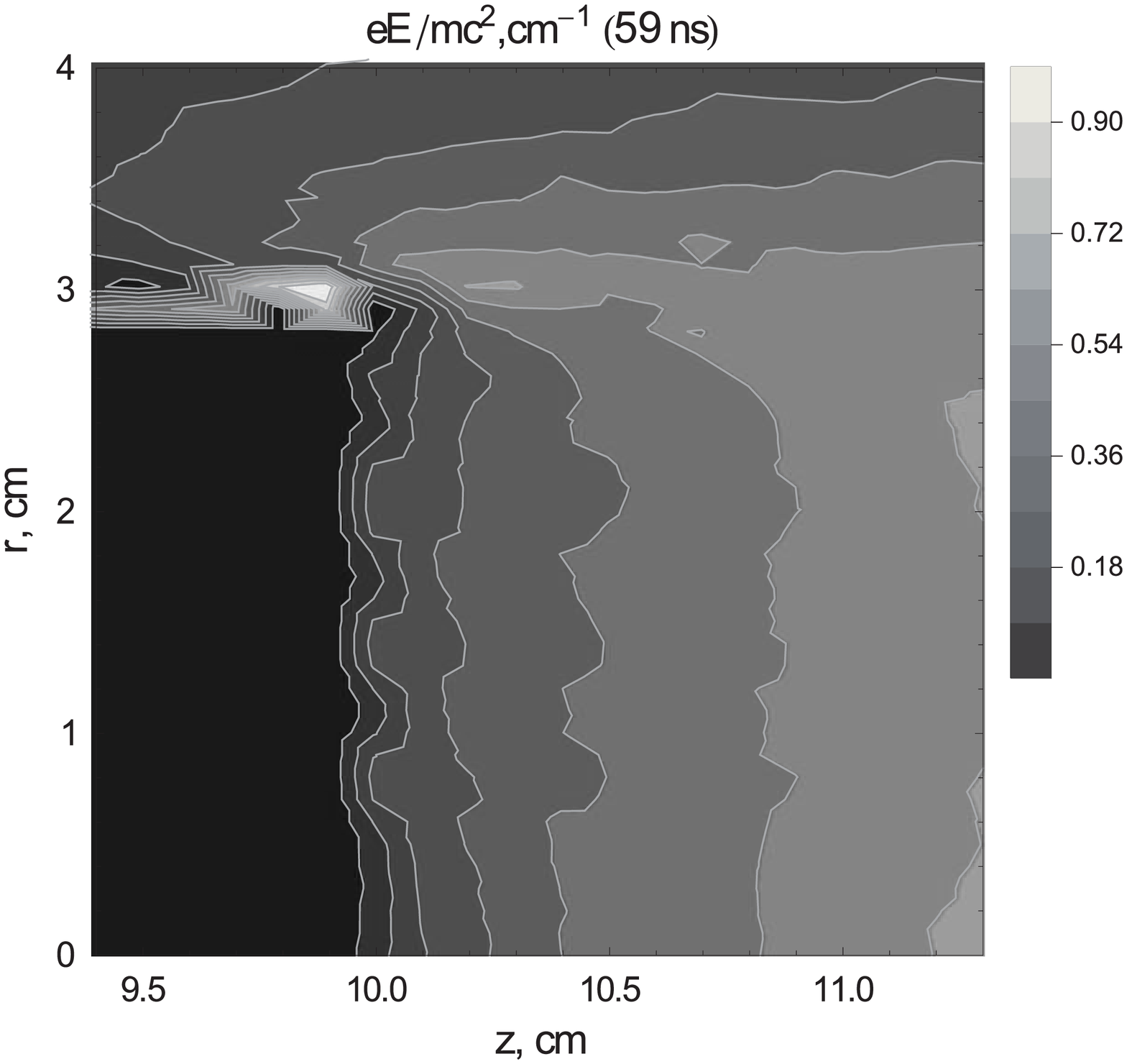}}\\
\end{center}
\caption{Electric field strength near the surface of the solid
cathode at 35ns, 40ns, 47ns and 59ns} \label{Fig.1add}
\end{figure}

\begin{figure}[ht]
\begin{center}
\resizebox{50mm}{!}{\includegraphics{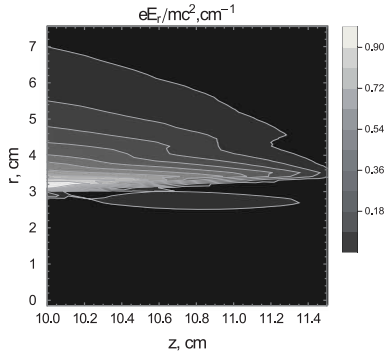}}\\
\end{center}
\caption{Radial electric field} \label{Fig.2add}
\end{figure}

\begin{figure}[ht]
\begin{center}
\resizebox{50mm}{!}{\includegraphics{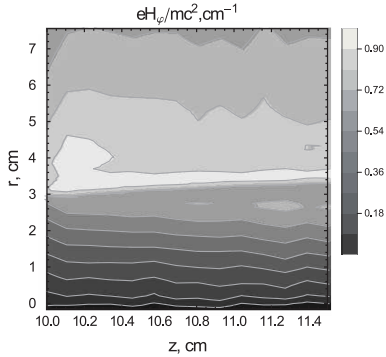}}\\
\end{center}
\caption{Magnetic field} \label{Fig.3add}
\end{figure}

Let us consider the typical behavior of the electric and  magnetic
fields illustrated by  a diode with the cathode-anode gap
$h=1.5$~cm and a solid cathode   of 3~cm radius. As is seen in
Figs. \ref{Fig.2add} and \ref{Fig.3add}, the radial component
$E_r$ of the electric field and the magnetic field $B_{\varphi}$
increase as the distance $r$ from the diode axis is increased
until $r$ becomes equal to the cathode radius, after which the
fields start decreasing. Let us note that substantially different
from zero values  of the electric-field radial component
responsible for the radial repulsion
 force are concentrated near the periphery of the cathode.
 The force
 $eE_r$ acting on the relativistic particles decreases appreciably as
 the particles come closer to the anode.
 This happens due to vanishing the tangential component of the
 electric field at the conducting surface (anode).

The  minimum in the momentum spread arises from the
combined effects of two competing processes: Coulomb repulsion
between charged particles and  pinching of the electron beam by
the self-induced magnetic field.
At small cathode radius, the radial Coulomb repulsion prevails,
being most intense near the external edge of the cathode, whereas
at large radius, the compression by the self-induced magnetic
field starts to dominate. Both processes lead imparting a radial
momentum to the beam electrons.
At a certain value of the radius, the Coulomb repulsion reduces
the effect produced by the self-induced magnetic field, and
pinching does not occur.
The electrons arriving at the anode have the same energies; the
increase in the radial velocity is accompanied by the decrease in
the longitudinal one.
%
%
Because different increments are imparted to the momenta of the
electrons escaping from the cathode at different distances from
the system axis, the observed momentum spread occurs.

For an infinite plane cathode with finite emitting surface
(Fig.\ref{fig:geometry}a), the effects of Coulomb repulsion
slacken (Fig.\ref{fig:1_spread}) because the radial component of
the electric field vanishes identically along the entire
conducting surface of the cathode. No additional sharpening occurs
in the electric field repulsing the particles. As a result,
radial motion of the beam is determined by the magnetic field
increasing as the radius of the emitting surface is increased.

\section{Current density distribution}

{The developed code also enables getting the time picture of beam
generation and evaluating the contribution coming from different
cathode regions to the produced current.}
The emission from both  solid and  ring-type cathodes starts
mainly from the cathode outer edge, and no difference between the
two cathode types is observed during the first $5 \div 10$~ns.
After a short time ($<50$~ns), the whole surface of the solid
cathode emits almost uniformly with some local current density
spikes approximately twice as high as the average current density
value.
The distance between these spikes correlates with the distance
between the cathode jets (these spikes are manifestation of
nonuniformity of expanding plasma surface discussed in previous
subsection).
For the ring-type cathode, the current distribution near the
cathode center  grows rapidly after the first 10~ns and remains
several times higher than the average current density value.

For both cathodes the beam diameter exceeds the cathode diameter
approximately 1.5 times.

\begin{figure}
  \centerline{\includegraphics[width=5.5 cm]{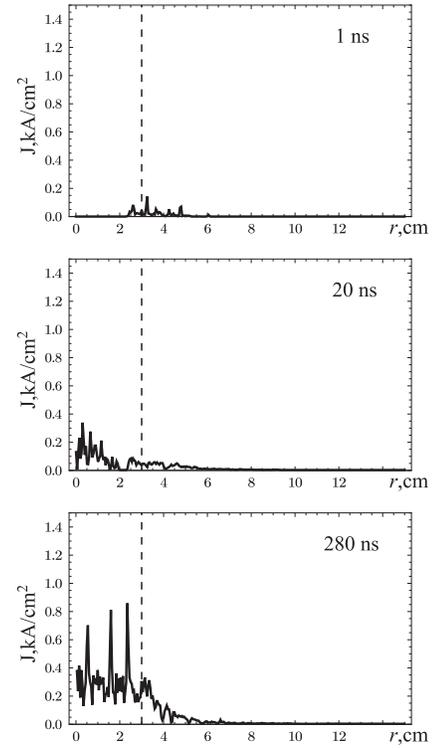}}
  \caption{Current density evolution for the solid cathode, dashed vertical line marks the cathode radius $r_c$}
  \label{fig:current_density_solid}
\end{figure}

\begin{figure}[t]
  \centerline{\includegraphics[width=5.5 cm]{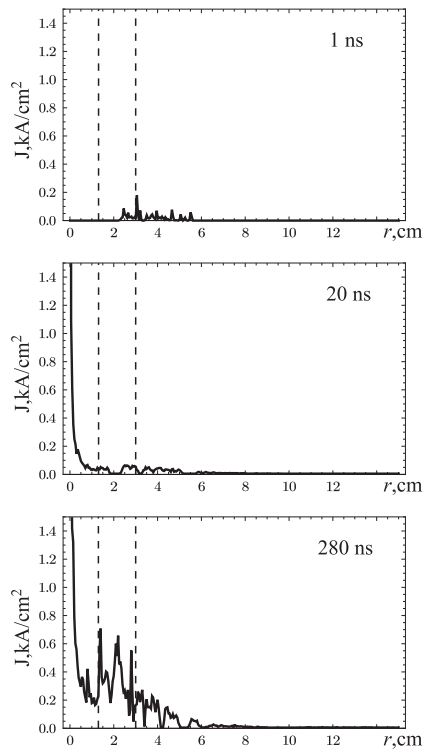}}
  \caption{Current density evolution for the ring-type cathode, dashed vertical lines mark the cathode
  outer $r_c$ and inner $r_{c1}$ radii}
  \label{fig:current_density_ring-type}
\end{figure}

The explosive electron emission is most intense from the nonplanar
regions of the cathode, particularly, from its inner edge. Due to
Coulomb repulsion, charged particles make their way to the region
that is not occupied by the beam. As a consequence, at the axis of
a relativistic vacuum diode, the density of a high-current beam
significantly exceeds the average current density in the
cathode-anode gap.

%
A similar mechanism leads to the formation of current-density
peaks (maxima)
(Figs.\ref{fig:current_density_solid},\ref{fig:current_density_ring-type}).
%
%
%
%
The electrons that escape from the convex explosive-emission
center are repulsed. The interaction between these repulsing
microflows gives rise to peaks and dips in the current density
distribution
(Figs.\ref{fig:current_density_solid},\ref{fig:current_density_ring-type}).

As the distance between emission centers on the cathode surface is
determined by risetime of voltage pulse applied to the cathode,
plasma expansion speed and some other factors. For the parameters
used for simulation (voltage pulse risetime 30ns, plasma expansion
speed 2~cm/$/mu$s) this distance is about 0.5-1cm. The same
distance the peaks in
Figs.\ref{fig:current_density_solid},\ref{fig:current_density_ring-type}
are spaced. Typical space mesh step and time-step are as small as
0.05~cm and 1.3~ns, respectively. Therefore, the peaks are
perfectly resolved.

\section{Conclusion}
The electron beam dynamics in a vacuum diode has been simulated.
The guidelines on how to match the size of the cathode (cathode
brim), the emitting area, and the cathode-anode gap for getting
the lowermost momentum spread in the produced electron beam have
been developed.
The distributions of current density over the beam radius obtained
for both solid and ring-type cathodes enable explaining the anode
mesh burning in the center in the case of a ring-type cathode.


\section*{Acknowledgment}

The authors would like to thank Prof. Vladimir Baryshevsky for
incentive and valuable discussions.

This research was funded by Ministry of Education and Science of
Russian Federation, grant no. 14.577.21.0092 project ID
RFMEFI57714X0092.



%

\end{document}